\documentclass[sigconf]{acmart}
\usepackage{multirow}
\usepackage{makecell}
\AtBeginDocument{%
  \providecommand\BibTeX{{%
    \normalfont B\kern-0.5em{\scshape i\kern-0.25em b}\kern-0.8em\TeX}}}

\acmSubmissionID{}

\begin{document}

\title{Reprogramming Self-supervised Learning-based Speech Representations for Speaker Anonymization}

\author{Xiaojiao Chen}
\affiliation{%
  \institution{Xinjiang University}
  \city{Urumqi}
  \country{China}
}\email{xiaojiaoch@163.com}

\author{Sheng Li}
\affiliation{%
  \institution{NICT}
  \city{Kyoto}
  \country{Japan}}
\email{sheng.li@nict.go.jp}

\author{Jiyi Li}
\affiliation{%
  \institution{University of Yamanashi}
  \city{Kofu}
  \country{Japan}
}\email{jyli@yamanashi.ac.jp}

\author{Hao Huang}
\affiliation{%
 \institution{Xinjiang University}
  \city{Urumqi}
  \country{China}}
\email{hwanghao@gmail.com}

\author{Yang Cao}
\affiliation{%
  \institution{Hokkaido University}
  \city{Sapporo}
  \country{Japan}}
\email{yang@ist.hokudai.ac.jp}

\author{Liang He}
\affiliation{%
  \institution{Tsinghua University}
  \city{Beijing}
  \country{China}}
\email{heliang@tsinghua.edu.cn}

\renewcommand{\shortauthors}{Chen, et al.}

\begin{abstract}
Current speaker anonymization methods, especially with self-supervised learning (SSL) models, require massive computational resources when hiding speaker identity. This paper proposes an effective and parameter-efficient speaker anonymization method based on recent End-to-End model reprogramming technology. To improve the anonymization performance, we first extract speaker representation from large SSL models as the speaker identifies. To hide the speaker's identity, we reprogram the speaker representation by adapting the speaker to a pseudo domain. Extensive experiments are carried out on the VoicePrivacy Challenge (VPC) 2022 datasets to demonstrate the effectiveness of our proposed parameter-efficient learning anonymization methods. Additionally, while achieving comparable performance with the VPC 2022 strong baseline 1.b, our approach consumes less computational resources during anonymization. 
\end{abstract}



\settopmatter{printacmref=false}

\maketitle

\section{Introduction}
\label{sec:intro}
Speech is an essential and rich source of personal information, including unique tract shape, accent, rhythm, or other sensitive attributes. As personal information widely applies on mobile phones, bank self-service voice systems, and forensic testing~\cite{singh2018forensic}, the risk of an attacker maliciously using such information increases. Considering recent security issues, it is impossible to ignore the existence of speaker privacy protection. Therefore, many efforts have been dedicated to protecting the speaker's privacy, and one of the main approaches is speaker anonymization. 

Speaker anonymization technology aims to suppress speaker identity information in the speech signal for speaker privacy. Recently, different approaches \cite{hashimoto2016privacy,fang2019speaker,srivastava2020design,turner2020speaker} have been proposed to hide the speaker's identity. Moreover, the emergence of x-vector \cite{snyder2018x} based approaches further flourishes the speaker anonymization task. The x-vector is chosen because it can effectively represent speaker identity in recognition systems. 

\citet{fang2019speaker}, for the first time, proposed an x-vector-based speaker anonymization framework. This framework has four main parts: feature extraction, x-vector anonymization, Mel-fbank generation, and speech synthesis.
They select the x-vectors within an external pool of speakers and average them to obtain a target pseudo-speaker x-vector in the x-vector anonymization part.
To improve the quality of anonymized speech, \citet{srivastava2022privacy} adopt HiFi-GAN \cite{kong2020hifi} to generate a high-quality anonymized waveform with generative adversarial (GAN) network. 
\citet{turner2020speaker} analyze the framework's shortcomings in \cite{fang2019speaker}, which generated x-vectors that tend to be much more similar to each other than the original x-vectors. Hence, they modify the framework in \cite{fang2019speaker} with a generative network to learn the distributional properties of the x-vector space.

\citet{perero2022x} also proposed an x-vector anonymization method by an autoencoder-adversarial network, which combines the autoencoders \cite{goodfellow2016deep} and domain adversarial training \cite{ganin2016domain}. Through this encoder-decoder network, they transform the source x-vector to a new pseudo-x-vector, where speaker identity has been hidden. Some researchers start with the data distribution of the vector. 

\citet{mawalim2020x} proposed two x-vector anonymization approaches based on singular value decomposition and statistical properties. They modify the singular value of the input x-vector, decompose the input x-vector based on its statistical properties, and transform it with regression models. 
Although these methods perform well, too many computing resources are required.

Self-supervised learning (SSL), which can learn robust speech representations by unlabeled data, has recently attracted wide attention in speech processing. SSL's success is because it can learn universal speech representations, which can be adapted effectively across various speech processing. Due to SSL's outstanding performance and generalization ability, an increasing number of SSL frameworks \cite{schneider2019wav2vec,baevski2020wav2vec,hsu2021hubert,chen2022wavlm} have been proposed and widely applied to various downstream tasks, such as automatic speech recognition (ASR), automatic speaker verification (ASV), and speech separation. 

Recent neural reprogramming technology only uses some small additive modules to adapt a frozen pre-trained model from a source domain to a target domain. It has been proposed as one parameter-efficient learning method for out-of-domain image prediction \cite{elsayedadversarial} and mitigating the gap between the different languages \cite{yang2021voice2series}.

This paper proposes an anonymization method based on reprogramming large SSL models to improve speaker anonymization performance and reduce computational resources. We extract the speech representations from HuBERT \cite{hsu2021hubert}, Wav2vec 2.0 \cite{baevski2020wav2vec}, and WavLM \cite{chen2022wavlm} respectively and use these speech representations as speaker identities. Then, we reprogram representation embedding to hide the speaker identity by adapting the speaker to a pseudo domain with a small account of training parameters. We carry out experiments on the test datasets of VPC 2022. Specifically, results on two datasets, VCTK and LibriSpeech, show that our method effectively hides speaker identity. Moreover, our method consumes less computational complexity than the baseline \cite{srivastava2022privacy}.

The main contributions of this paper are as follows: (1) We propose a novel speaker anonymization method by reprogramming the speaker representation to hide the speaker's identity. To our knowledge, this is the first study in which speaker anonymization is applied by reprogramming. (2) The reprogramming-based speaker anonymization framework is based on the current state-of-the-art SSL pre-trained models. 

\section{Proposed Method}
\begin{figure}[tp]
    \centering
	{\includegraphics[scale=0.375]{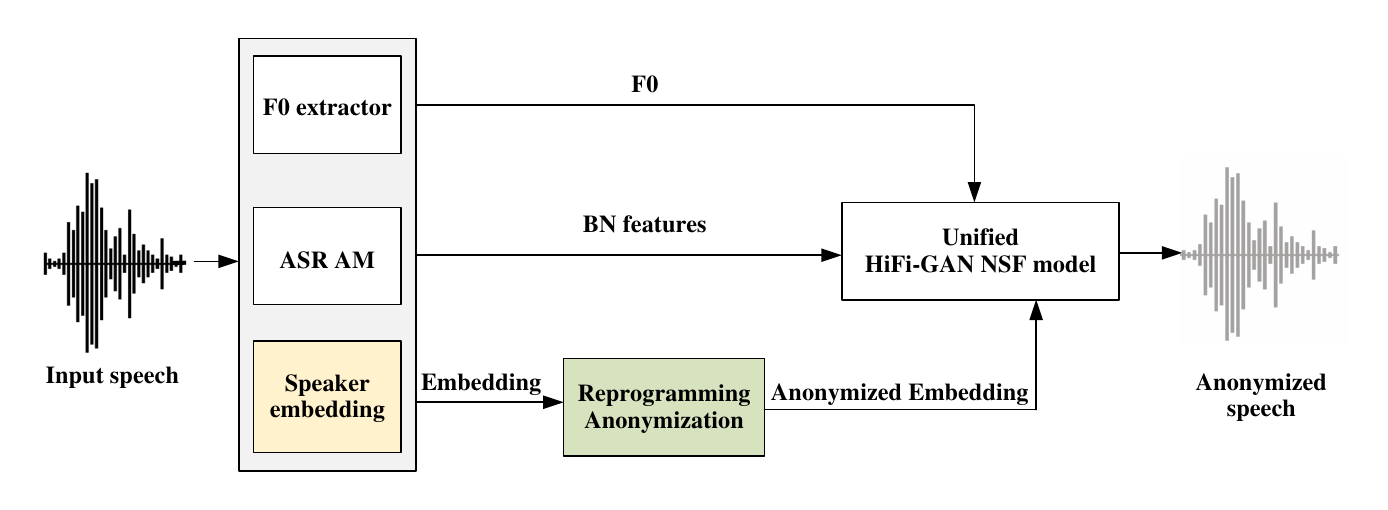}}
	\caption{Architecture of parameter-efficient anonymization method. The parameter of the pre-trained (white and yellow) model is frozen, and the anonymization part (green) is a trainable reprogramming for hiding speaker identity.}
	\label{fig:noise}
\end{figure}

Our proposed reprogramming anonymization method is illustrated in Fig.\ref{fig:noise}, which is based on the framework in \cite{srivastava2022privacy}. There are three steps in the x-vector-based framework: feature extraction, reprogramming anonymization, and speech synthesis. The feature extraction step prepares the pith (F0) and bottleneck (BN) features as speech content representation from ASR and the speaker embedding extracted from large SSL models. To hide the input speech's source speaker identity, embedding is reprogrammed by adapting the source speaker to a pseudo-speaker domain. To obtain a high-quality anonymized speech, NSF model \cite{wang2019neural} and HiFi-GAN \cite{kong2020hifi} are combined to synthesize anonymized speech. 

\subsection{Motivation}
In the baseline, the anonymization method based on an x-vector generates an anonymized x-vector by averaging candidate x-vectors from an external x-vector pool of speakers. Given a source speaker, probabilistic linear discriminant analysis (PLDA) \cite{ioffe2006probabilistic}, and cosine distance is used as a distance measure between candidate vectors and the x-vector of the source speaker. 
For x-vectors, $\omega_i$ of a source speaker and x-vectors $\omega_j$ of a candidate speaker, the cosine distance between them is: $d_{cosine}(\omega_i,\omega_j) =1-\frac{{\omega_i} \cdot {\omega_j}}{||\omega_i||_{2}||\omega_j||_{2}}$. Then, rank all the x-vectors in the candidate pool in increasing order of their distance from a source speaker and select either the top $K$ (near) or the bottom $K$ (far). The anonymized x-vector of the source speaker is the mean of selected $K$ candidate vectors. 
The anonymity of a speaker often requires a calculation of the speaker distance of $M$ times and a weighted sum of $K$ selected candidate speakers, where $M$ is the number of candidate x-vectors and $M>>K$.
It is seen that such a great anonymity effect of the baseline method comes at the cost of a lot of computation and a large external x-vector pool. Motivated by the weakness of the baseline, which needs a large computational source, we aim to improve the anonymized performance and reduce the computational source in two steps. 
First, we extract the speaker embedding from self-supervised pre-trained models. Then, we propose a reprogramming anonymization method by reprogramming a source speaker's representation without any external speaker pool and with less time and parameters. 

\subsection{Proposed Reprogramming Anonymization}
\label{adv}
The large self-supervised model requires massive labeled and unlabeled data in a pretraining and fine-tuning stage. The model parameters also learn the massive speaker information and have achieved great success in ASV tasks than the x-vector. Therefore, to obtain greater speaker representation and improve speaker anonymization, we extract speech representation as speaker identity from three well-known self-supervised pre-trained models, including Wav2vec 2.0 \cite{baevski2020wav2vec}, HuBERT \cite{hsu2021hubert}, and WavLM \cite{chen2022wavlm}. Given a pre-trained SSL model $G(\cdot)$ and a pre-trained speech synthesis model $f(\cdot)$. We can synthesize the speaker's waveform $Z=f(X,BN,F0)$ by feeding the input to the pre-trained $f(\cdot)$. Source speaker $i$ representation $X_{ij}$ is expressed as: $X_{ij} = G(x_{ij})$, where the $x_{ij}$ is $j$th input speech of speaker $i$. Fundamental frequency ($F0_{ij}$) and bottleneck ($BN_{ij}$) features of speaker $i$ can be extracted from speech $x_{ij}$. 

To ensure that all trial utterances from a given speaker are uttered by the same pseudo-speaker while different pseudo-speakers utter trial utterances from different speakers. We anonymize the source speaker at the speaker level. We pre-process $X_{ij}$ to a normalized speaker-level representation $X_{i}$ of speaker $i$: $X_{i} = \frac{1}{N}\sum_{j=1}^{N} X_{ij}$,
where $N$ is the total number of speaker $i$'s utterances. Note that dimension of $X_{i}$ is $1 \times 512$. Then we reprogram $X_{i}$ by frozen the pre-trained $G(\cdot)$ and $f(\cdot)$. Let the reprogrammed speaker representation $X_{i}^{'}$ can be formulated as:
$X_{i}^{'}= X_i + W \odot M$,
where $W_{1 \times 512}$ is a set of trainable parameters for reprogramming speaker identity, and $M_{1 \times 512}$ is a binary masking matrix that indicates the location of $X_i$ needs training. The $\odot$ operator denotes element-wise
product. 
In previous reprogramming studies \cite{yang2021voice2series,huck2023english,yen2021study}, $W \odot M$ can be seen as a trainable additive input transformation for reprogramming. In this paper, let $\theta=W \odot M$ with dimension of ${1 \times 512}$. And the $\theta$ is trained to reprogram the speaker identity. From the calculated complexity perspective, our proposed method needs less computation when generating the pseudo-speaker. 

For anonymization, the farther distance between $X_{i}^{'}$ and $X_i$, the better the anonymization performance is.
We consider the mean absolute error (MAE) to constrain the size of speaker embedding reprogramming. Thus, the whole anonymization method can be reformulated as follows:
\begin{equation}\small
    \theta^{*} =\mathop{ \arg\min}\limits_{\theta} {{\mathcal L}(f(X_{i}^{'},BN_{ij},F0_{ij}),Z)- ||X_{i}^{'} - X_i||},
\end{equation}
where $\mathcal L$ is the speech synthesis model's loss function, note that the speech synthesis model's parameters are frozen, and only the parameters $\theta$ are trainable. Obtaining the $\theta$ can adapt the $X_i\in {\mathbb{R}^{S}}$ to a pseudo domain $X_{i}^{'}\in {\mathbb{R}^{P}}$ and $\theta$ can make anonymized speaker $X_{i}^{'}$ as a valid input to the speech synthesis model $f(\cdot)$. 

\subsection{Minimum reprogramming distance}
However, in our empirical study, the current constraint of reprogramming speaker identify $||X_{i}^{'} - X_i||$ affects the quality of the synthesized anonymized speech and may cause overfitting. We further impose stricter restrictions on reprogramming speaker representations, which can be expressed as:
\begin{equation}\scriptsize
    \begin{aligned}
    \theta^{*} =\mathop{ \arg\min}\limits_{\theta} {{\mathcal L}(f(X_{i}^{'},BN_{ij},F0_{ij}),Z)- ||X_{i}^{'} - X_i||} 
    s.t.\ ||X_{i}^{'} - X_i||< \epsilon ,
    \end{aligned}
\end{equation}
where $\epsilon$ is a weight norm penalty and a small value to trade off the anonymization performance and usability of anonymized speech. 

\begin{table*}
\footnotesize
  \begin{center}
  \renewcommand{\arraystretch}{1.2}
  \caption{EER($\%$): Anonymized performance results for test partitions (o-original, a-anonymized speech) on the $ASV_{eval}$.}
  \label{tab:eer1}
    \begin{tabular}{cc|cc|cc|cc|cc|cc|cc}
    \toprule
     \multicolumn{2}{c|}{}& \multicolumn{4}{c|}{LibriSpeech}& \multicolumn{4}{c|}{VCTK-different}& \multicolumn{4}{c}{VCTK-common} \\
     \multicolumn{2}{c|}{Gender} & \multicolumn{2}{c|}{Female} &\multicolumn{2}{c|}{Male} & \multicolumn{2}{c|}{Female} &\multicolumn{2}{c|}{Male}& \multicolumn{2}{c|}{Female} &\multicolumn{2}{c}{Male}\\
     \multicolumn{2}{c|}{Anony.Type} & OA & AA& OA & AA& OA & AA& OA & AA& OA & AA& OA & AA\\
     \midrule
     \multicolumn{2}{c|}{Baseline} & \textbf{51.64}& 27.55 & \textbf{52.12} & 34.97 & \textbf{49.64} & 22.69 & \textbf{55.11} & 25.03 & \textbf{51.73} & 22.25 & \textbf{53.67} & 24.58\\
     \multirow{3}{*}{\makecell[c]{Speaker\\Embedding} } 
     &HuBERT Base & 45.44 & 37.59 & 47.22 & 46.33 &\textbf{49.07} & 35.39 & \textbf{49.25} &37.72 & 46.53 & 37.28  &47.74  & 39.27\\
     &Wav2vec2 Base & \textbf{46.17} & 39.23 & 48.78 & \textbf{46.55} & \textbf{49.49} & 39.76 & 48.34 & \textbf{ 42.31} & 45.38 & 37.86 & 46.61 & \textbf{39.55}\\
     & WavLM Base & \textbf{46.17} & \textbf{39.42} & \textbf{50.11}& 46.33 & 48.82 & \textbf{40.53} & 48.45 & 41.91 & 47.11 & \textbf{39.88} & 45.76 & 37.29\\
    \bottomrule
    \end{tabular}
  \end{center}
\end{table*}

\begin{table*}
\footnotesize
  \begin{center}
  \renewcommand{\arraystretch}{1.2}
  \caption{EER($\%$): Anonymized performance results for test partitions (o-original, a-anonymized speech) on the $ASV^{anon}_{eval}$.}
  \label{tab:eer2}
    \begin{tabular}{cc|cc|cc|cc|cc|cc|cc}
    \toprule
     \multicolumn{2}{c|}{}& \multicolumn{4}{c|}{LibriSpeech}& \multicolumn{4}{c|}{VCTK-different}& \multicolumn{4}{c}{VCTK-common} \\
     \multicolumn{2}{c|}{Gender} & \multicolumn{2}{c|}{Female} &\multicolumn{2}{c|}{Male} & \multicolumn{2}{c|}{Female} &\multicolumn{2}{c|}{Male}& \multicolumn{2}{c|}{Female} &\multicolumn{2}{c}{Male}\\
     \multicolumn{2}{c|}{Anony.Type} & OA & AA& OA & AA& OA & AA& OA & AA& OA & AA& OA & AA\\
     \midrule
    \multicolumn{2}{c|}{Baseline} & \textbf{50.55} & 9.49 & \textbf{48.33}& 7.80 & \textbf{52.99} & 10.91 &52.70 & 7.52 & \textbf{53.18} & 15.32 & \textbf{50.85} & 8.19\\
     \multirow{3}{*}{\makecell[c]{Speaker\\Embedding} }& 
     HuBERT Base & 41.06 & \textbf{18.07} & 42.09 & 13.81 & 39.92  &20.52  &47.70  &16.36  & 41.04 &18.79 & 44.63 & 14.69\\
     & Wav2vec2 Base & \textbf{48.36} & 16.06 & 
     \textbf{45.66} & \textbf{14.70} & \textbf{46.55} & \textbf{26.13} & \textbf{53.21} & 13.83 & \textbf{46.82} & \textbf{25.66} & \textbf{50.56} & \textbf{17.23} \\
    &WavLM Base & 39.23 & 15.51 & 39.87 & 13.36 & 39.04 &20.32 &47.30 & \textbf{19.58} & 41.04 & 19.65 & 42.94 & 14.97\\
    \bottomrule
    \end{tabular}
  \end{center}
\end{table*}

\begin{table}[h]
\footnotesize
  \begin{center}
  \renewcommand{\arraystretch}{1.2}
  \caption{WER\% achieved by $ASR^{anon}_{eval}$ on data processed by Baseline and our proposed method vs. WER achieved by $ASR_{eval}$ on the original (Orig.) unprocessed data.}
  \label{tab:wer}
    \begin{tabular}{lc|cc|cc}
      \toprule
       \multicolumn{2}{c|}{}& \multicolumn{2}{c|}{Libri.} & \multicolumn{2}{c}{VCTK}  \\
      \multicolumn{2}{c|}{Anonymization system} & Dev. & Test & Dev. & Test\\
      \midrule
       \multicolumn{2}{c|}{Orig.} & 3.82 & 4.15 & 10.79 & 12.82\\
       \multicolumn{2}{c|}{Baseline} & 4.19 & 4.43 & 10.98 & 10.69\\
       \multirow{3}{*}{\makecell[c]{Speaker\\Embedding}}& HuBERT Base & 6.00 & 6.05 & 18.35  &17.49 \\
       & Wav2vec2 Base& 5.81   &5.53  & 16.81 & 16.32 \\
       & WavLM Base&5.72  &5.84 &14.48  &14.20  \\
      \bottomrule
    \end{tabular}
  \end{center}
\end{table}

\section{Experimental Setup}
\label{}
\noindent\textbf{Datasets}:
Recently, researchers have further standardized the system evaluation criteria for speaker anonymity systems \cite{tomashenko2020introducing,tomashenko2022voiceprivacy}. Therefore, most datasets used in this experiment were from the VoicePrivacy 2022 Challenge\cite{tomashenko2022voiceprivacy}. We also compared our proposed approaches with baseline 1.b in \cite{tomashenko2022voiceprivacy}. The LibriSpeech train-clean-100 was used to train the speech synthesis system. After obtaining a trained speech synthesis system, train-clean-100 was used to train $\theta$. Moreover, the train-clean-360 of LibriSpeech was used to train the anonymized automatic speaker verification ($ASR_{eval}$) and ASR systems ($ASV_{eval}$). The ASR is a factorized time delay neural network (TDNN-F) model architecture \cite{povey2018semi} and is trained using the Kaldi toolkit \cite{povey2011kaldi}. An x-vector extractor with a TDNN model topology is also trained using Kaldi.
We also anonymized the train-clean-360 of LibriSpeech to train $ASR^{anon}_{eval}$ and $ASV^{anon}_{eval}$. Moreover, the development and test sets comprise LibriSpeech dev-clean and a subset of the VCTK corpus to evaluate $ASR_{eval}$, $ASR^{anon}_{eval}$, $ASV_{eval}$ and $ASV^{anon}_{eval}$. 

\noindent\textbf{Experimental Setup}:
Speaker representation is respectively extracted from three pre-trained SSL models, including Wav2vec 2.0 \cite{baevski2020wav2vec}, HuBERT \cite{hsu2021hubert}, and WavLM \cite{chen2022wavlm}. They are trained by LibriSpeech 960h. The weights of all pre-trained SSL models, which we directly extract the speech representation, are obtained from SUPERB \cite{yang2021superb}.
The output embedding of these three models is with a dimension of (batch\_size, sequence\_length, hidden\_size), and the hidden\_state = 768. We added feed-forward networks before the speech synthesis system to transfer dimensions 768 to 512 for subsequent dimension matching. We initialize the $W \odot M$ and optimize 512 parameters of $W \odot M$ with the LibriSpeech train-clean-100. And during training, we set the $\epsilon=0.1$. The speech synthesis model in this paper is the same as \cite{tomashenko2022voiceprivacy}, which combines the NSF model as the generator with the discriminators of HiFi-GAN \cite{kong2020hifi} and can directly convert BN, F0, and x-vector features using an NSF mode. To balance the quality of synthesized anonymous audio with the anonymization effect, we fine-tuned the synthesis system separately with different speaker representations using LibriSpeech train-clean-100. After training, the discriminators can be safely discarded, and only the trained NSF is used in the anonymization system.

\noindent\textbf{Attack Scenarios}:
For the evaluation, attackers were assumed to have access to the source speech and anonymized speech utterances. In this paper, we mainly consider two attack scenarios: (1) Ignorant attacker: (OA: \textbf{O}rignal enrollment-\textbf{A}nonymized trial): One or more anonymized trial utterances are exposed to the attacker and use an ASV system (denoted $ASV_{eval}$). (2) Lazy-informed (AA: \textbf{A}nonymized enrollment-\textbf{A}nonymized trial): The attacker has access to the original enrollment data and the anonymization system. The attacker generates the speaker-level anonymization of the enrollment data. However, the trial and enrollment data are anonymized using different pseudo-speakers, which causes the mismatch between trial and enrollment. The attacker can retrain an ASV system (denoted $ASV^{anon}_{eval}$).

\noindent\textbf{Evaluation metrics}: Two main metrics are employed. They are the equal error rate (EER) as the privacy metric and the word error rate (WER) as the primary utility metric. EER is a privacy metric that assumes the attacker has access to one trial and several enrollment utterances. WER is to evaluate how well the anonymization method can preserve linguistic information.

\section{Experiment Results}
\label{results}
\subsection{Anonymized performance}

We first evaluate the anonymized speakers by the ASV systems, which can directly show whether the speaker identity is successfully hidden. Moreover, we utilize the equal error rate (EER\%) for evaluation (a larger value means anonymization works). 
Table \ref{tab:eer1} and Table \ref{tab:eer2} show the results of anonymized performance, measured by $ASV_{eval}$ and $ASV^{anon}_{eval}$ on the test set in ignorant attacker (OA) and lazy-informed (AA) scenarios, respectively. We also evaluate our proposed reprogramming anonymization method with three different speakers embedding extracted from the pre-trained SSL model.
Our goal under both scenarios is to achieve a high anonymization performance because the higher the EER is, the better anonymization.

From Table \ref{tab:eer1}, our proposed method leads to a notable increase in lazy-informed (AA) conditions compared with the baseline. The results demonstrate our proposed method can effectively hide the speaker's identity.
For the ignorant attacker (OA) condition, the proposed reprogramming anonymization method can be comparable or effectively reduce the distance between the proposed one and the baseline. 
Results in Table \ref{tab:eer2} demonstrate the great anonymization performance of the proposed reprogramming method in AA condition. Compared to the baseline, the proposed method with three speakers embedding all achieves a competitive score OA condition with a much lower computation cost.

Moreover, we identified two interesting findings from Table \ref{tab:eer1} and \ref{tab:eer2}. Firstly, we discovered that the outcomes produced by the different conditions were more similar than significantly different from those generated by the baseline system. This finding indicates that our reprogramming anonymization method effectively reduces the difference between the OA and AA conditions. Secondly, our proposed approaches demonstrated a similar anonymization performance across different datasets and speaker embeddings. As the speaker embeddings are extracted from SSL models, the robust speech representation contributes to the anonymization performance and may explain the phenomenon as mentioned above. 

\subsection{Intelligibility evaluation results}

ASR is used to assess the ability of the anonymization system to preserve linguistic information. Table \ref{tab:wer} shows the objective intelligibility evaluation in terms of the word error rate (WER) in the ASR evaluation system($ASR^{anon}_{eval}$),  trained by the anonymized utterance. 
There is 2\% of WER distance by our SSL-based system and baseline, but transitional language models or other post-error correction methods can easily correct it. This sacrifice is acceptable. We hypothesize that speaker embeddings from SSL models may be susceptible to interference from external factors, causing a lack of robustness and affecting the utility of anonymized speech.

\subsection{Speaker embedding}
We evaluated the performance of the reprogramming anonymization method using different speaker embeddings extracted from three pre-trained SSL models. Our results, as shown in Tables 1 and 2, indicate that the anonymization performance is similar across all three speaker embeddings, and all of them outperform the baseline under AA conditions. Notably, the speaker embedding extracted from Wav2vec 2.0 demonstrates the best anonymization performance. By balancing the anonymity effect and the ASR effect, our proposed method balances anonymous synthesized audio quality and the anonymity effect.

\section{Conclusions}
This paper proposes a novel speaker anonymization method called the reprogramming anonymization method. Motivated by a large amount of computation for the baseline, the reprogramming method is a parameter-efficient anonymization method, which only needs to train 512 parameters during anonymization. We explore the speaker representation by learning from three different SSL models, and they effectively reduce the anonymity effect in different conditions.
The experimental results demonstrated that our proposed adversarial anonymization method can outperform the baseline in lazy-informed conditions. Moreover, our proposed method uses fewer computational sources. In future work, we will focus on improving the quality of anonymized speech that better suits the anonymization systems.

\bibliographystyle{ACM-Reference-Format}
\bibliography{mybib}

\end{document}